\documentclass[iop]{emulateapj}
\usepackage{epsfig}
\usepackage{apjfonts}
\usepackage{color}
\usepackage{aas_macros}

\begin{document}

\title{Black hole central engine for ultra-long gamma-ray burst 111209A and its associated supernova 2011kl}
\author{He Gao\altaffilmark{1}, Wei-Hua Lei\altaffilmark{2}, Zhi-Qiang You\altaffilmark{1} and Wei Xie\altaffilmark{2}}

\altaffiltext{1}{Department of Astronomy, Beijing Normal University, Beijing 100875, China}
\altaffiltext{2}{School of Physics, Huazhong University of Science and Technology,
Wuhan, 430074, China.}
\email{gaohe@bnu.edu.cn; leiwh@hust.edu.cn}

\begin{abstract}
Recently, the first association between an ultra-long gamma-ray burst (GRB) and a supernova is reported, i.e., GRB 111209A/SN 2011kl, which enables us to investigate the physics of central engines or even progenitors for ultra-long GRBs. In this paper, we inspect the broad-band data of GRB 111209A/SN 2011kl. The late-time X-ray lightcurve exhibits a GRB 121027A-like fall-back bump, suggesting a black hole central engine. We thus propose a collapsar model with fall-back accretion for GRB 111209A/SN 2011kl. The required model parameters, such as the total mass and radius of the progenitor star, suggest that the progenitor of GRB 111209A is more likely a Wolf-Rayet star instead of blue supergiant, and the central engine of this ultra-long burst is a black hole. The implications of our results is discussed.
\end{abstract}

\keywords{accretion, accretion disks - black hole physics - gamma-ray burst: individual (GRB 111209A) }

%%%%%%%%%%%%%%%%%%%%%%%%%%%%%%%%%%%%%%%%
\section{Introduction}

Recently, ``ultra-long bursts'', a subclass of gamma-ray bursts (GRBs) with unusually long central engine activity ($\sim$ hours)  compared to typical GRBs (tens of seconds), have been paid great attention \citep{gendre13,virgili13,stratta13,levan14,greiner15}. In some references, the ``ultra-long" GRBs refered to the GRBs with $\gamma$-ray duration $T_{90}$ extending to $\sim 10^3$ s or even larger\footnote{Although no clear boundary line has been defined yet.} \citep{gendre13,virgili13,stratta13,levan14}. Some authors \citep{zhang14,gao15}, on the other hand, argue that $T_{90}$ is not a reliable measure for defining ``ultra-long" GRBs, taking into account the prolonged central engine activity time of some GRBs, indicated by flares \citep{burrows05a,zhang06,margutti11} and shallow decay plateaus \citep{troja07,liang07} in the early X-ray afterglow light curves. They propose to redefine the burst duration as $t_{\rm burst}$ by taking into account both $\gamma$-ray and the aforementioned X-ray light curve features.  Based on both observational analysis and numerical simulations, they found that bursts with duration $T_{90}$ of order $10^3$ s can be reproduced with a normal central engine, while bursts with $t_{\rm burst} > 10^4$ s require an extended central engine activity, may be classified to the ``ultra-long'' population \citep{zhang14,gao15}.

Even though lack of a clear definition, it is generally considered that the ultra-long GRBs may have either a special central engine or a special progenitor \citep{levan14}. For instance, some authors \citep{gendre13,nakauchi13,levan14} proposed a blue supergiant-like progenitor for ultra-long GRBs \citep{meszaros01,nakauchi13}, considering their much larger radii could naturally explain the unusually long durations. On the other hand, it is also proposed that the ultra-long GRBs may have a special central engine, such as a strongly magnetized millisecond neutron star (a magnetar) \citep{levan14,greiner15}. Research on the physical origin of ultra-long GRBs would potentially promote our understanding of the central engine and progenitor of GRBs.

Most recently, \cite{greiner15} reported the first discovered association between an ultra-long GRB and a supernova, i.e., GRB 111209A/SN 2011kl. Based on the observed properties of SN 2011kl, such as its spectra and light curve shape, they rule out a blue supergiant progenitor and a tidal disruption interpretation for GRB 111209A. Nevertheless, based on the unexpected high luminosity of SN 2011kl, which is intermediate between canonical overluminous GRB-associated supernovae and super-luminous supernovae \citep{galyam12,quimby11}, they suggest that $^{56}$Ni is not responsible for powering the luminosity of SN 2011kl,  but an additional energy input is required. They thus propose that the central engine of GRB 111209A might be a millisecond magnetar.

Considering the comprehensive observations of GRB 111209A/SN 2011kl, especially that the late-time X-ray light curve of GRB 111209A exhibits a GRB 121027A-like fall-back bump \citep{wu13}, rather than obey the dipole radiation profile, the central engine of GRB 111209A is more like a black hole (BH) instead of a magnetar. The question is could the BH central engine serve as the energy reservoir to power the abnormally luminous SN 2011kl ? In this work, we intend to interpret the broadband data of GRB 111209A/SN 2011kl within the collapsar model: the GRB central engine is a BH; a fraction of the materials in the envelope would fall back and reactive the accretion onto the BH, tapping the spin energy of the BH to give rise the unusually long central engine activity timescale; the energy and angular momentum could also be extracted magnetically from the revived accretion disc, depositing energy into the supernova ejecta to give rise the unusually high luminosity of SN 2011kl. In section 2, we describe the observational features of GRB 111209A. A general picture of the fall-back accretion model is given in section 3 and we apply this model to the broadband data of GRB 111209A/SN 2011kl in section 4. In section 5, we briefly summarize our results and discuss the implication.

%%%%%%%%%%%%%%%%%%%%%%%%%%%%%%%%%%%%%%%%
\section{Observational features of GRB 111209A}

GRB 111209A was discovered at $T_0$ = 2011:12:09-07:12:08 UT on 9 December 2011 by the Burst Alert Telescope (BAT) on board {\it Swift}, and was later accurately located by XRT at a position of RA(J2000)=00$^h$ 57$^m$ 22.63$^s$ and Dec(J2000)=-46d 48$'$ 03.8$''$, with an estimated uncertainty of 0.5 arcsec \citep{hov11} . GRB 111209A showed an extraordinarily long prompt duration, and was monitored up to $T_0+1400$ s until BAT entering the orbital gap region. It was also detected by the Konus detector on the WIND spacecraft \citep{golenetskii11}. As shown in the ground data analysis of the Konus-Wind instrument, GRB 111209A was recorded with a continuous coverage extending from 5,400 s before to 10,000 s after the Swift trigger $T_0$.

Both Very Large Telescope (VLT)/X-shooter (2011 December 10 at 1:00 UT) and Gemini-N/GMOS-N detected the early spectroscopy of the transient optical light of GRB 111209A. A redshift of $z = 0.677$  was suggested by the identify of absorption lines and emission lines from the host galaxy \citep{vreeswijk11}. Based on the Konus-Wind results, GRB 111209A had a fluence of $(4.86 \pm 0.61) \times 10^{-4}$ erg cm$^{-2}$ \citep{golenetskii11}, inferring an isotropic gamma-ray energy release $E_{\gamma,\rm iso}=4\pi D_{L}^2 f_{\gamma}/(1+z) = 5.54 \pm 0.70 \times10^{53}$ erg. Here we adopt the concordance cosmology with $\Omega_m=0.27$, $\Omega_{\Lambda}=0.73$ and $h_0=0.71$.

{\it Swift}/XRT observations started at 425 s after the BAT trigger \citep{hov11}, revealing a bright afterglow mainly with several components:
\begin{itemize}
\item An initial plateau phase overlapping with the prompt $\gamma$-ray emission;
\item After an observational gap, from $3\times 10^3$ s to $\sim 10^4$ s, the light curve rapidly rise back and then  gradually steepen to a ``steep decay" phase with decay index of $\sim 5$, behaving very likely the GRB 121027A-like fall-back bump \citep{wu13}.
\item At very late time ($\sim 10^5$ s), the tail of steep decay phase is superposed by another power-law component with decay index of $\sim 1.5$, which is usually denoted as ``normal decay" phase in the canonical picture of X-ray afterglow \citep{zhang06}.
\end{itemize}

The afterglow of GRB 111209A was also clearly detected in the optical-UV band by {\it Swift}/UVOT and other ground based instruments, such as the TAROT-La Silla \citep{klotz11} and the seven-channel optcial/near-infrared imager GROND \citep{kann11}. After the prompt optical flashes, the earlier optical afterglow shows a normal power-law decay until day 15. And then, the optical light curve starts to deviate from the power-law decay and remains essentially flat between days 15 and 30. After day 30, the light curve starts to decay again, approaching the host-galaxy level. Most recently, based on its temporal and spectral features, this excess emission is identified as a supernova, designated SN 2011kl, associated with GRB 111209A \citep{greiner15}. The bolometric peak luminosity of SN 2011kl is intermediate between canonical over-luminous GRB-associated supernovae and super-luminous supernovae.

%%%%%%%%%%%%%%%%%%%%%%%%%%%%%%%%%%%%%%%%

\section{Fall-back accretion model}

In this work, we intend to use the collapsar model to interpret the broadband afterglow data of GRB 111209A. The physical picture is as follows: the progenitor star has a core-envelope structure, as is common in stellar models. The bulk of the mass in the core part collapses into a rapidly spinning black hole (BH), and the rest mass forms a surrounding accretion disc. The GRB prompt emission can be powered by the Blandford-Znajek (1977, hereafter BZ) mechanism, in which the spin energy of the BH is extracted via the open field lines penetrating the event horizon. An alternative mechanism for powering GRB jet is the neutrino annihilation process, which is too ``dirty'' and inffective to account for long-term activity of GRB, like ultra-long GRBs (Fan et al. 2005; Lei et al. 2009, 2013; Liu et al. 2015). The energy and angular momentum could also be extracted magnetically from accretions discs, by field lines that leave the disc surface and extend to large distance, centrifugally launching a baryon-rich wide wind/outflow through the Blandford-Payne (Blandford \& Payne 1982, hereafter BP) mechanism. The bounding shock responsible for the associated supernova and the BP outflow would transfer kinetic energy to the envelope materials. During the fall-back, only a portion of the fall-back mass accretes onto the BH, while the rest is ejected in a disc wind \citep{kumar08a,kumar08b}.
The more energetic the supernova shock, the less envelope material falls back into the center. For a parcel of gas of the progenitor star at radius $r_{\rm fb}$, the fall-back time could be estimated as $t_{\rm fb} \sim (\pi^2 r_{\rm fb}^3 /8 G M_\bullet)^{1/2}$, where $M_\bullet$ is the BH mass. The fall-back of the envelope materials may form a new accretion disc, powering the shallow decay phase or late flares seen in X-ray afterglow. The BP outflow from the new disc would further deposit energy into the supernova ejecta.

Based on some analytical and numerical calculations, the fall-back accretion rate initially increases with time as $\dot{M}_{\rm early} \propto t^{1/2}$ until it reaches a peak value at $t_{\rm p}$ \citep{macfadyen01,zhang08,dai12}. And then the late-time fall-back accretion behavior would follows $\dot{M_{\rm late}}\propto t^{-5/3}$, as suggested by \cite{chevalier89} until time $t_e$, when most of the fall back materials fuel out and the accretion behavior could be described with $\dot{M_{\rm final}}\propto t^{-\alpha}$, where $\alpha$ depends on the stellar structure and rotation rate of progenitor star\footnote{In this work, the value of $\alpha$ is inferred from the X-ray afterglow steep decay index.}. We use a three-segments-broken-power-law function of time to describe the evolution of the fall-back accretion rate as
\begin{eqnarray}
\dot{M} = &&\dot{M}_{\rm p} \left[ \frac{1}{2}\left(\frac{t-t_0}{t_{\rm p}-t_0} \right)^{-1/2} +  \frac{1}{2}\left(\frac{t-t_0}{t_{\rm p}-t_0} \right)^{5/3} \right]^{-1} \nonumber \\
&& \times\left[1+\left(\frac{t-t_0}{t_{\rm b}-t_0} \right)^{\alpha -5/3}\right]^{-1},
\label{dotm}
\end{eqnarray}
where $t_0$ is the starting time of fall-back accretion (henceforth, time is defined in the cosmologically local frame).

Consider a Kerr black hole with mass $M_\bullet$ and (or dimensionless mass $m_{\bullet}=M_{\bullet}/M_\odot$) and angular momentum $J_\bullet$. The BZ jet power is \citep{lee00,li00,wang02,mckinney05,lei08,lei11,lei13}
\begin{equation}
L_{\rm BZ}=1.7 \times 10^{50} a_{\bullet}^2 m_{\bullet}^2
B_{\bullet,15}^2 F(a_{\bullet}) \ {\rm erg \ s^{-1}},
\label{eq_Lmag}
\end{equation}
where $a_\bullet = J_\bullet c/(GM_\bullet^2)$ is the BH spin parameter, and $F(a_{\bullet})=[(1+q^2)/q^2][(q+1/q) \arctan q-1]$ with $q= a_{\bullet} /(1+\sqrt{1-a^2_{\bullet}})$. $B_{\bullet}$ is the magnetic field strength threading the BH horizon. Since the magnetic field on the BH is supported by the surrounding disc, one can estimate its value by equating the magnetic pressure on the horizon to the ram pressure of the accretion flow at its inner edge \cite[e.g.][]{moderski97},
\begin{equation}
\frac{B_{\bullet}^2}{8\pi} = P_{\rm ram} \sim \rho c^2 \sim \frac{\dot{M} c}{4\pi r_{\bullet}^2}
\label{Bmdot}
\end{equation}
where $r_{\bullet}=(1+\sqrt{1-a_\bullet^2})r_{\rm g}$ is the radius of the BH horizon, and $r_{\rm g} = G M_\bullet /c^2$. We can then rewrite the BZ power as a function of mass accretion rate as
\begin{equation}
L_{\rm BZ}=9.3 \times 10^{53} a_\bullet^2 \dot{m}  X(a_\bullet) \ {\rm erg \ s^{-1}} ,
\label{eq:EB}
\end{equation}
and
\begin{equation}
X(a_\bullet)=F(a_\bullet)/(1+\sqrt{1-a_\bullet^2})^2.
\end{equation}

The the observed X-ray luminosity is connected to the BZ power via the X-ray radiation efficiency $\eta$ and the jet beaming factor $f_{\rm b}$, i.e.,
\begin{equation}
\eta L_{\rm BZ}=f_{\rm b} L_{\rm X,iso}
\label{eq1}
\end{equation}

The BP outflow luminosity could be estimated as \citep{AN99}
\begin{eqnarray}
L_{\rm BP}=\frac{(B_{\rm ms}^{\rm P})^2r_{ms}^4\Omega_{ms}^{2}}{32c}
\end{eqnarray}
where $B_{\rm ms}^{\rm P}$ and $\Omega_{ms}$ are the poloidal disk field and the Keplerian angular velocity at inner stable circular orbit radius ($r_{ms}$).
Here we define $R_{\rm ms} = r_{\rm ms}/r_{\rm g}$ as the radius of the marginally stable orbit in terms of $r_{\rm g}$. The expression for $R_{\rm ms}$ is \citep{bardeen72},
\begin{eqnarray}
R_{\rm ms} =  3+Z_2 -\left[(3-Z_1)(3+Z_1+2Z_2)\right]^{1/2},
\end{eqnarray}
for $0\leq a_{\bullet} \leq 1$, where $Z_1 \equiv 1+(1-a_{\bullet}^2)^{1/3} [(1+a_{\bullet})^{1/3}+(1-a_{\bullet})^{1/3}]$, $Z_2\equiv (3a_{\bullet}^2+Z_1^2)^{1/2}$. The quantity $\Omega_{ms}$ is given by
\begin{eqnarray}
\Omega_{ms}=\frac{1}{M_\bullet/c^3(\chi^{3}_{ms}+a_{\bullet})}
\end{eqnarray}
where $\chi_{\rm ms}\equiv \sqrt{r_{\rm ms}/{M_\bullet}}$. Following Blandford \& Payne 1982, $B_{\rm ms}^{\rm P}$ could be estimated as
\begin{eqnarray}
B_{\rm ms}^{\rm P}=B_{\bullet}(r_{ms}/r_H)^{-5/4}
\end{eqnarray}
where $r_{H}=M_\bullet(1+\sqrt{1-a_{\bullet}^2})$ is the horizon radius of the black hole.

The photospheric luminosity of SNe powered by a variety of energy sources (e.g. radioactive $^{56}$Ni decay, BP power injection, etc) could be expressed as \citep{arnett82,wang15a,wang15b}
\begin{eqnarray}
L(t)&=&\frac{2}{\tau_{m}}e^{-\left(\frac{t^{2}}{\tau_{m}^{2}}+\frac{2R_{0}t}{v\tau_{m}^{2}}\right)}~
\left(1-e^{-\tau_{\gamma}(t)}\right)\int_0^t e^{\left(\frac{t'^{2}}{\tau_{m}^{2}}+\frac{2R_{0}t'}{v\tau_{m}^{2}}\right)}   \nonumber\\
     &&\times\left(\frac{R_{0}}{v\tau_{m}}+\frac{t'}{\tau_{m}}\right)L_{\rm inj}(t')dt'~\mbox{erg s}^{-1},
\label{equ:lum}
\end{eqnarray}
where $R_{0}$ is the initial radius of the progenitor, which could be taken as zero to largely simplify the above equation since it is very small compared to the radius of the ejecta. $\tau_{m}$ is the effective light curve timescale, which reads
\begin{eqnarray}
\tau_{m}=\left(\frac{2\kappa M_{\rm ej}}{\beta vc}\right)^{1/2},
\label{equ:tau_m}
\end{eqnarray}
where $\kappa$, $M_{\rm ej}$ and $v$ are the Thomson electron scattering opacity,
the ejecta mass, and the expansion velocity of the ejecta. $\beta \simeq 13.8$ is a constant that accounts for the density distribution of the ejecta \citep{wang15a,wang15b}. $L_{\rm inj}$ is the generalized energy source. $\left(1-e^{-\tau_{\gamma}(t)}\right)$ reflects the $\gamma$-ray trapping rate, with
 $\tau_{\gamma}(t)~=~At^{-2}$ being the optical depth to $\gamma$-rays \citep{cha09,cha12}. Provided that the SN ejecta has a uniform density distribution  ($M_{\rm ej}~=~(4/3) \pi \rho R^3$, $E_{\rm K}~=~(3/10)M_{\rm ej}v^2$),
the characteristic parameter $A$ could be estimated as
\begin{eqnarray}
A&=&\frac{3\kappa_{\gamma} M_{\rm ej}}{4\pi v^2 }\nonumber \\
 &=&4.75 \times 10^{13}\left(\frac{\kappa_{\gamma}}{0.1~{\rm cm}^2~{\rm g}^{-1}}\right)  \nonumber \\
 &&\times\left(\frac{M_{\rm ej}}{M_\odot}\right)\left(\frac{v}{10^9~{\rm cm}~{\rm s}^{-1}}\right)^{-2}~\mbox{s}^{2},
\label{equ:leakage}
\end{eqnarray}
where $\kappa_{\gamma}$ is the opacity to $\gamma$-rays.

As argued in \cite{greiner15}, radioactive $^{56}$Ni decay could not be responsible for the luminosity of SN 2011kl. Here we take that
\begin{eqnarray}
L_{\rm inj}\simeq L_{\rm BP}.
\end{eqnarray}

\begin{figure*}[t]
\begin{center}
\begin{tabular}{ll}
\resizebox{80mm}{!}{\includegraphics[]{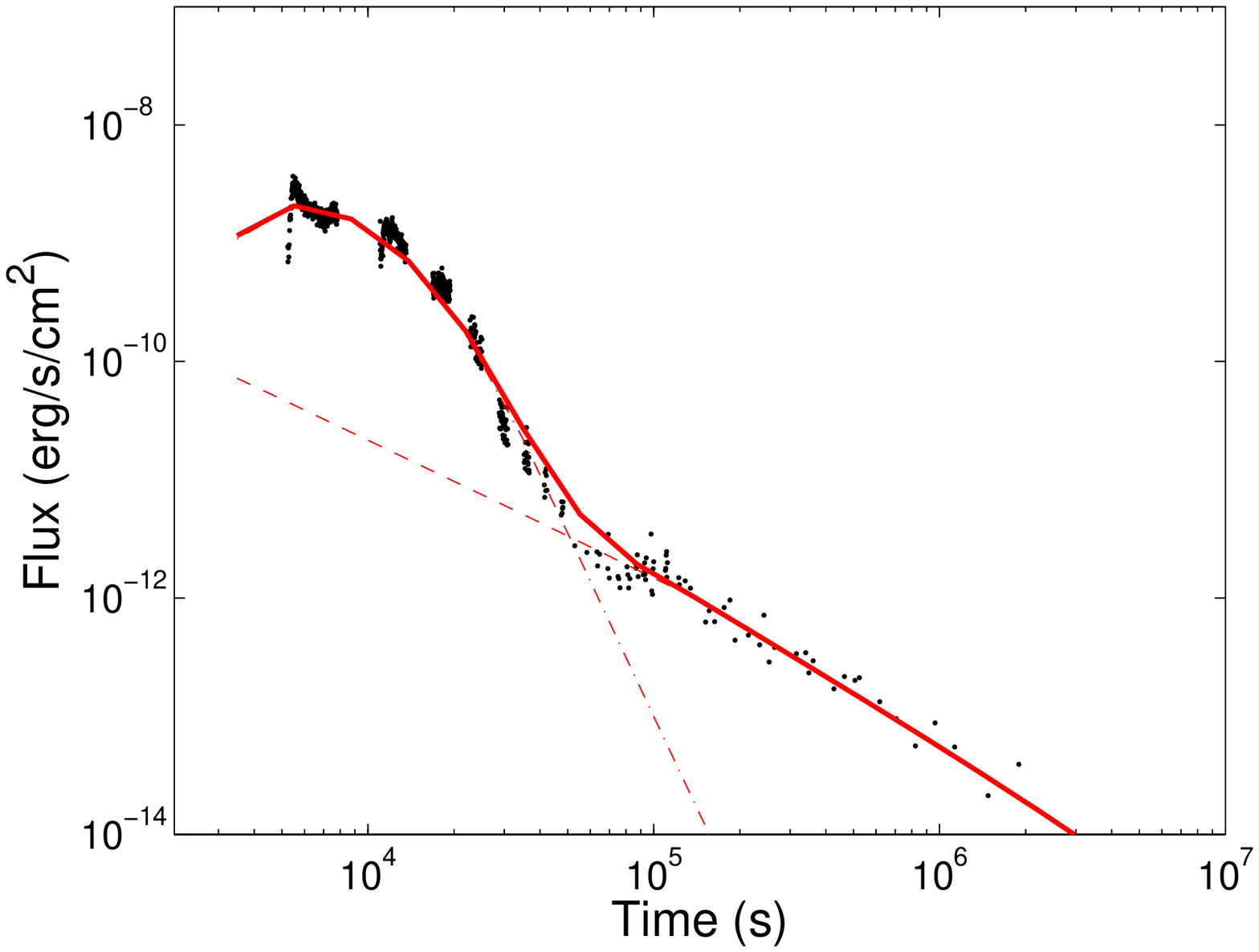}} &
\resizebox{80mm}{!}{\includegraphics[]{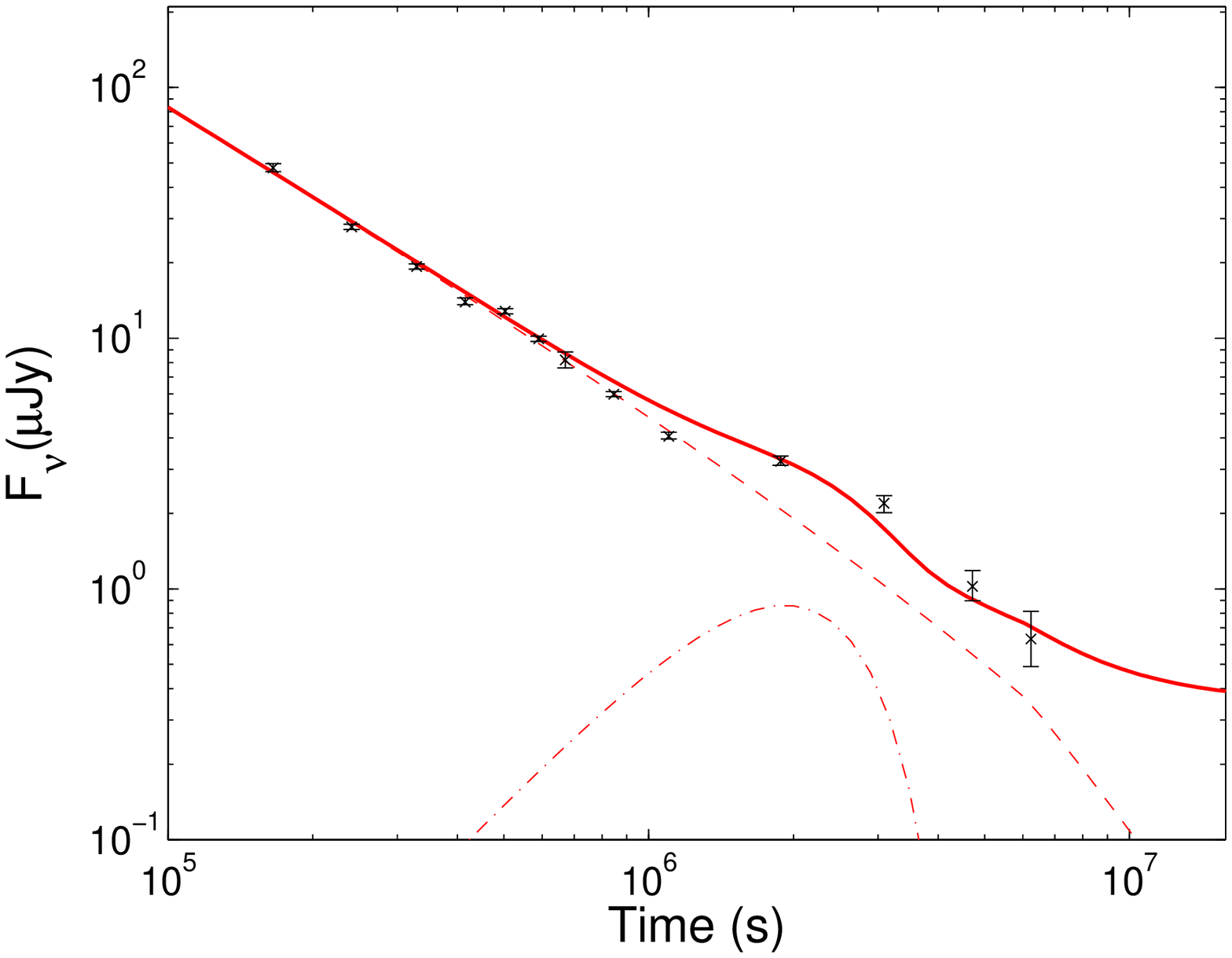}} \\
\end{tabular}
\caption{Modeling results for the XRT range (left panel) and r band (right panel) light curve of GRB 111209A.
The observed data are exhibited with points, and the theoretical modeling are shown with solid lines.
The thin-dash line denotes the external shock component and the dash-dotted line in left (right) panel denotes the BZ powered radiation according to the fall-back disc (the BP outflow powered supernova component). For optical data, a constant host galaxy emission is invoked \citep{greiner15}.}
\label{fig:fit}
\end{center}
\end{figure*}

%%%%%%%%%%%%%%%%%%%%%%%%%%%%%%%%%%%%%%%%
\section{Application to GRB 111209A}
In the following, we apply the above fall-back accretion model to fit the broad-band data of GRB 111209A, as presented in Figure \ref{fig:fit}. Table 1 summarize the values of the parameters adopted in the model. We find that the broadband data of GRB 111209A could be well explained with all standard parameter values.

The early X-ray light curve (from $3000$s to $5\times10^{4}$s) could be well interpreted with the BZ power induced by the fall-back disc. We focus on the overall shape of the X-ray lightcurve. The flare-like variations during this phase could be due to the fragmentation during the fall-back phase \citep{king05}.
The contribution from the GRB afterglow emission is initially outshone by the BZ power and emerge later ($>5\times10^{4}$s) to account for the late time normal decay light curve.  As shown in Figure \ref{fig:fit}, the early optical data may come from the GRB afterglow emission, while the late optical data, i.e., the SN 2011kl component, could be dominated by the emission from a BP process powered supernova.

In the interpretation for the afterglow component, the jet isotropic kinetic energy $E_k$ of  $7.6\times10^{52}~{\rm erg}$ and the ambient medium density $n$ of $1~{\rm cm^{-3}}$ are adopted\footnote{It is worth noting that the fitness of the GRB afterglow emission could be also achieved by simultaneously enhancing the kinetic energy but reducing the medium density, which means $E_k$ could be equal or larger than $10^{53}~{\rm erg}$  \citep{nakauchi13}, matching with the isotropic gamma-ray energy release $E_{\gamma,\rm iso}$ in order of magnitude.} . The fitting results depend weakly on the values of initial Lorentz factor ($\Gamma_0$) and half opening angle ($\theta$) of the jet.
The microphysics shock parameters (e.g., $\epsilon_e$, $\epsilon_B$, and $p$) are all chosen as their commonly used values in GRB afterglow modeling. \cite[][for a review]{gao13,kumarzhang15}. To compare with the observations of X-ray bump in GRB 111209A, we carried out numerical calculation for the time evolution of the BZ power. The radiation efficiency $\eta=0.05$ is taken in our calculation. The BH is initially set up with a mass $m_\bullet=3$ and a spin $a_\bullet=0.9$. The fall-back accretion starts at $t_0=3000/(1+z)$ s, peaks at $t_p=8000/(1+z)$ s and fuels out around $t_b=1.8\times10^{4}/(1+z)$ s. The outermost radius of the fallback material could be essentially estimated as $r_{\rm fb}\sim (8 G M_\bullet t_b^2/\pi^2)^{1/3}=3.2\times 10^{11} $ cm, which is smaller than the typical radius of blue supergiant star \citep{nakauchi13}. To achieve the fitness of the data, the late time accretion rate decay index $\alpha=5$ is required. For the supernova ejecta, we take the standard values for its mass ($M_{\rm ej}\sim3{\rm M_{\odot}}$), initial velocity ($v_i=0.06c$) and the effective opacity $\kappa=0.06~{\rm cm^{2}~g^{-1}}$ \citep{lyman16}. Note that these ejecta parameters suffer severe degeneracy, which could be justified by equation \ref{equ:tau_m}.

From Eqs.(3) and (6), the maximum angular velocity and magnetic field strength around BH could be estimated as
\begin{equation}
\Omega_{\bullet}= \frac{c^3}{G M_\bullet} \frac{a_\bullet}{2 (1+\sqrt{1-a_\bullet^2})} \simeq 1.01 \times 10^{5}  m_\bullet^{-1}  q \ \rm rad/s.
\end{equation}
and
\begin{equation}
B_{\bullet, \rm p} \simeq 2.4 \times 10^{14} L_{\rm X,iso,49}^{1/2} m_\bullet^{-1} q a_\bullet^{-2} X^{-1/2}(a_\bullet) \eta_{-2}^{-1/2} f_{\rm b,-2}^{1/2} \rm G.
\end{equation}
With initial setup values, e.g., $m_\bullet=3$ and a spin $a_\bullet=0.9$, we have $B_{\bullet, \rm p}\sim 10^{14}$G and $\Omega_{\bullet}\sim2.1\times10^{4}~\rm rad/s$. In this case, the BH spin period is around 0.3 ms. 

From Eqs. (9) and (10), the initial angular velocity and magnetic field at $r_{\rm ms}$ could be estimated as
\begin{equation}
\Omega_{\rm ms} \simeq 2.03 \times 10^{5}  m_\bullet^{-1}  (\chi_{ms}^3 + a_\bullet)^{-1} \ \rm rad/s.
\end{equation}
and
\begin{equation}
B_{\rm ms, \rm p} \simeq 2.4 \times 10^{14} L_{\rm X,iso,49}^{1/2} (r_{ms}/r_H)^{-5/4} m_\bullet^{-1} q a_\bullet^{-2} X^{-1/2}(a_\bullet) \eta_{-2}^{-1/2} f_{\rm b,-2}^{1/2} \rm G.
\end{equation}
With initial setup values, we have $B_{\rm ms, \rm p}\sim 0.5\times10^{14}$G and $\Omega_{\rm ms}\sim1.5\times10^{4}~\rm rad/s$.

\begin{table*}
\begin{center}{\scriptsize
\caption{Parameters for interpreting the broadband data of GRB\,111209A.}
\begin{tabular}{ccccccc} \hline\hline
 \multicolumn{7}{c}{BH and ejecta parameters}\\
  \hline
& $M_\bullet~({\rm M_{\odot}})$                    & $a_\bullet$    &$M_{\rm ej}~({\rm M_{\odot}})$                    &  $v/c$   & $\kappa~({\rm cm^{2}~g^{-1}})$     \\
&$3$     &  $0.9$         & $3$     &  $0.06$         &$0.06$\\
   \hline
  \multicolumn{7}{c}{GRB afterglow parameters}\\
  \hline
$E_k~({\rm erg})$     &$\Gamma_0$               &  $n~(\rm{cm^{-3}})$& $\theta~({\rm rad})$&$\epsilon_e$ &  $\epsilon_B$&  $p$  \\
$7.6\times 10^{52}$     &  $200$ &  $1$   & $0.4$& $0.1$     &  $10^{-4}$&  $2.5$        \\
  \hline
  \multicolumn{7}{c}{Other parameters}\\
  \hline
 $\eta$& $t_0~({\rm s})$ & $t_p~({\rm s})$ & $t_b~({\rm s})$&$\dot{M}_p~({\rm M_{\odot}}s^{-1})$&$\alpha$&$\kappa_{\gamma}~({\rm cm^{2}~g^{-1}})$\\
$0.05$  & $3000/(1+z)$ &   $8000/(1+z)$ &   $1.8\times10^{4}/(1+z)$ &$2.5\times10^{-4}$ &$5$&$0.1$   \\
   \hline\hline
 \end{tabular}
 }
\end{center}
\end{table*}

\section{Conclusion and discussion}

As the first reported association between ultra-long GRB and supernova, GRB 111209A/SN 2011kl system provides us a good chance to study the properties of central engines or even progenitors for ultra-long GRBs. In this work, we apply a fallback accretion scenario within the collapsar model to interpret the broadband data of GRB 111209A/SN 2011kl. We find that with all standard parameter values, both X-ray and optical observations could be well explained. In our interpretation, the central BH mass, the fallback material mass and the supernova ejecta mass are adopted as $3~M_\odot$, $\sim 2.6~M_\odot$ and $3~M_\odot$ respectively, inferring that the total mass of the progenitor star is in order of $10~M_\odot$. Moreover, the outermost radius of the fallback material is estimated as $r_{\rm fb}\sim 3.2\times 10^{11} $ cm, 5 times of solar radius. Therefore, we suggest that the progenitor of GRB 111209A is more likely a Wolf-Rayet star instead of blue supergiant star, which is consistent with the observational implications from \cite{greiner15}, although we argue that the central engine of this ultra-long burst is a BH rather than a magnetar. The required magnetic field strength around BH is $\sim10^{14}$ G, similar to the magnetar magnetic field properties ( $6-9\times10^{14}$ G) as proposed by \cite{greiner15}. But the black hole spin period is around 0.3 ms, almost 2 orders of magnitude faster than the assumed magnetar spin ($\sim12$ ms). 

If our interpretation is correct, the following implications can be inferred.

Under the framework of collapsar model, the central engine (BH) activity timescale could have a wide range, depending not only on the size of the progenitor star, but also on the stellar structure and rotation rate of the progenitor star \citep{kumar08a,kumar08b}. The latter property would mainly affect the fallback process of the envelope material, which could largely extend the central engine activity time, having chance to give rise to the ultra-long GRBs. On the other hand, the bounding shock responsible for the associated supernova and the BP outflow from the initial accretion disc would transfer kinetic energy to the envelope materials. If the injected kinetic energy is less than the potential energy of the envelop material, the starting time of the fallback would be delayed, which may even prolong the burst duration. However, if the injected kinetic energy is larger, which might be the majority cases, the fallback process is vanished and the central engine activity is relatively short, corresponding to the normal long GRBs. It is worth noting that besides GRB 111209A, the other GRB that exhibits a fall-back bump in X-ray light curve, namely GRB 121027A, is also an ultra-long burst \citep{wu13}.

Regardless of the fallback process, BP outflow would always deliver additional energy to the supernova ejecta, which may explain the fact that GRB associated supernovae is usually a energetic hypernovae \cite[][for details]{li16}. For ultra-long GRBs, where fallback accretion might perform, the associated supernovae could be even brighter, like SN 2011kl, since the BP outflow injection is largely prolonged. It is worth noticing that given the ejecta properties, such as its radial density profile, metallicity content and so on, the supernova spectrum is essentially determined by the total injected energy from the central engine, no matter the central engine is magnetar or black hole. 
For the case of SN 2011kl, \cite{greiner15} have carefully reproduced its spectrum within the magnetar central engine scenario and interpreted the observations. With the parameters presented in Table 1, it is easy to show that around the SN peak time ($10^6$ s), the total injected energy from the BP outflow is similar to the magnetar injection with parameters adopted in \cite{greiner15}.  We thus claim that the SN spectrum from our model is expected to be similar with the results calculated in \cite{greiner15}.

During the fallback, the disk accretion rate  ($\dot{M_p}=2.5 \times 10^{-4} M_\sun/s$) is too low to ignite significant neutrino emission (Popham et al. 1999; Chen \& Beloborodov 2007). As a concequence, the accretion flow could be dominated by advection at this stage, i.e., an ADAF, which has strong mass outflow due to its positive Bernoulli constant. Therefore, not all of the fallback mass accrete onto the black hole. In contrast, the majority are ejected in the disk wind (Kumar et al. 2008). MacFadyen \& Woosley (1999) found that energy dissipation in the disk can launch a wind with significant $^{56}$Ni. The investigating by Li et al. (2016) indicate that large $^{56}$Ni mass are also produced in the magnetic outflow. This may give rise to the $^{56}$Ni mass observed in the nebular spectra of hypernovae through Fe emission (e.g. Mazzali et al 2001).

Numerical simulations suggested that the BZ mechanism can power highly collimated GRB jets. However, additional simulations will be needed to see whether the magnetically-driven disc wind could drive a more isotropic hypernova blast.

\acknowledgments

We thank the referee for the helpful comments which have helped us to improve the presentation
of the paper. This work is supported by the National Basic Research Program ('973' Program) of China (grants 2014CB845800), the National Natural Science Foundation of China under grants 11543005, U1431124, 11361140349 (China-Israel jointed program).

\end{document}